\documentclass[10pt]{article}
\usepackage[reqno]{amsmath}
\usepackage{epsfig}
\usepackage{array}
\usepackage{float}

\begin{document}


\def\ltap{\ \raisebox{-.4ex}{\rlap{$\sim$}} \raisebox{.4ex}{$<$}\ }
\def\gtap{\ \raisebox{-.4ex}{\rlap{$\sim$}} \raisebox{.4ex}{$>$}\ }
\newcommand{\deltaatm}{\mbox{$\Delta  m^2_{\mathrm{atm}} \ $}}
\newcommand{\deltaatmmax}{\mbox{$(\Delta  m^2_{\mathrm{atm}})_{ \! \mbox{}_{\mathrm{MAX}}} \ $}}
\newcommand{\deltaatmmin}{\mbox{$(\Delta  m^2_{\mathrm{atm}})_{ \! \mbox{}_{\mathrm{MIN}}} \ $}}
\newcommand{\deltasol}{\mbox{$ \Delta  m^2_{\odot} \ $}}
\newcommand{\deltasolmax}{\mbox{$ (\Delta  m^2_{\odot})_{ \! \mbox{}_{\mathrm{MAX}}} \ $}}
\newcommand{\deltasolmin}{\mbox{$(\Delta  m^2_{\odot})_{ \! \mbox{}_{\mathrm{MIN}}}  \ $}}
\newcommand{\utre}{\mbox{$|U_{\mathrm{e} 3}|$}}
\newcommand{\utremax}{\mbox{$|U_{\mathrm{e} 3}|^2$ }}
\newcommand{\utremin}{\mbox{$|U_{\mathrm{e} 3}|^2$ }}
\newcommand{\utretilda}{\mbox{ $\widehat{|U_{\mathrm{e} 3}|^2}$}}
\newcommand{\uuno}{\mbox{$|U_{\mathrm{e} 1}|^2$}}
\newcommand{\uunomax}{\mbox{$|U_{\mathrm{e} 1}|^2$ }}
\newcommand{\uunomin}{\mbox{$|U_{\mathrm{e} 1}|^2$ }}
\newcommand{\betabeta}{\mbox{$(\beta \beta)_{0 \nu}  $}}
\newcommand{\mefff}{\mbox{$ < \! m  \! > $}}
\newcommand{\meff}{\mbox{$\left|  < \! m  \! > \right| \ $}}
\newcommand{\hbeta}{$\mbox{}^3 {\rm H}$ $\beta$-decay \ }
\newcommand{\eV}{\mbox{$ \  \mathrm{eV} \ $}}
\newcommand{\deltatre}{\mbox{$ \ \Delta m^2_{32} \ $}}
\newcommand{\deltadue}{\mbox{$ \ \Delta m^2_{21} \ $}}
\newcommand{\ueuno}{\mbox{$ \ |U_{\mathrm{e} 1}|^2 \ $}}
\newcommand{\uedue}{\mbox{$ \ |U_{\mathrm{e} 2}|^2 \ $}}
\newcommand{\uetre}{\mbox{$ \ |U_{\mathrm{e} 3}|^2  \ $}}
\renewcommand{\thefootnote}{\alph{footnote}}


\hyphenation{par-ti-cu-lar}
\hyphenation{ex-pe-ri-men-tal}
\hyphenation{dif-fe-rent}
\hyphenation{bet-we-en}
\hyphenation{mo-du-lus}

\renewenvironment{thebibliography}[1]
	{\begin{list}{\arabic{enumi}.}
	{\usecounter{enumi}\setlength{\parsep}{0pt}
	 \setlength{\itemsep}{0pt} 
         \settowidth
	{\labelwidth}{#1.}\sloppy}}{\end{list}}

\topsep=0in\parsep=0in\itemsep=0in

\newcounter{arabiclistc}
\newenvironment{arabiclist}
	{\setcounter{arabiclistc}{0}
	 \begin{list}{\arabic{arabiclistc}}
	{\usecounter{arabiclistc}
	 \setlength{\parsep}{0pt}
	 \setlength{\itemsep}{0pt}}}{\end{list}}

\flushright{Ref. SISSA 87/2001/EP}
\flushright{November 2001}
\vskip 0.6cm

\centerline{{ \bf \large  Majorana Neutrinos, CP-Violation,
 Neutrinoless Double-Beta and }}
\centerline{{ \bf \large Tritium-Beta Decays }}

\begin{center}
\vspace{0.3cm} 
S. Pascoli$^{(a,b)}$~\footnote{To be published 
in the Proceedings of the Conference
 NANP'01, III International Conference on Non-Accelerator
New Physics, Dubna (Russia),  June 
19-23 
2001. 
}, 
~~S. T. Petcov$^{(a,b)}$~\footnote{Also at: Institute of Nuclear Research and
Nuclear Energy, Bulgarian Academy of Sciences, 1784 Sofia, Bulgaria}

\vspace{0.2cm}   
{\em $^{(a)}$ Scuola Internazionale Superiore di Studi Avanzati, 
I-34014 Trieste, Italy\\
}
\vspace{0.2cm}   
{\em $^{(b)}$ Istituto Nazionale di Fisica Nucleare, 
Sezione di Trieste, I-34014 Trieste, Italy\\
}

\end{center}

\vspace{2mm}

\abstract{
If the present or upcoming searches for 
neutrinoless double $\beta-$ (\betabeta-) decay 
give a positive result, the Majorana nature
of massive neutrinos will be established.
From  the determination of the value of the
\betabeta-decay
effective Majorana mass parameter (\meff),
it would be possible to obtain 
information on the type of 
neutrino mass spectrum.
Assuming 3-$\nu$ mixing and 
massive Majorana neutrinos,
we discuss the information
a measurement of, or an upper bound 
on, \meff
can provide on the value of the lightest 
neutrino mass $m_1$. With additional 
data on the 
neutrino masses 
obtained in   $^3$H $\beta-$decay experiments,
it might be possible to 
establish whether the CP-symmetry is
violated in the lepton sector.
This would require very high precision measurements.
If CP-invariance holds, the allowed patterns of
the relative CP-parities
of the massive Majorana neutrinos would be 
determined.}

\normalsize\baselineskip=15pt


\renewcommand{\baselinestretch}{0.7}

\section{Introduction}
\vspace{-0.2cm}
  
   With the accumulation of more and stronger 
evidences for oscillations of the
atmospheric\cite{SKatm00}
and solar\cite{SKYSuz00SNO} neutrinos,  
caused by  neutrino mixing 
(see, e.g., \cite{BiPe87}), 
the problem of the nature of massive neutrinos 
emerges as one of the fundamental problems 
in the studies of neutrino mixing. 
Massive neutrinos can be Dirac or Majorana particles.
In the former case they possess 
a conserved lepton charge and
distinctive antiparticles, 
while in the latter there is no conserved 
lepton charge 
and massive neutrinos  are truly neutral 
particles identical with their antiparticles
(see, e.g., \cite{BiPe87}).
Thus, the question of the nature of 
massive neutrinos is directly related 
to the question of the 
basic symmetries of the 
fundamental particle interactions.

The present and upcoming 
neutrino oscillation experiments 
will allow to make a big step forward
in  understanding  the 
patterns of neutrino mass squared differences and 
of $\nu$-mixing, but will not be able to determine 
the absolute values of the neutrino masses and 
to answer the question regarding 
the nature of massive neutrinos.
The  \hbeta experiments, studying the electron spectrum, 
are sensitive to the electron (anti-)neutrino mass  
$m_{\nu_e}$ and can give information  on 
 the absolute value of neutrino masses.
The present bounds on $m_{\nu_e}$ from 
the Troitzk~\cite{MoscowH3} and Mainz~\cite{Mainz} 
 experiments read (at 95\%~C.L.): $m_{\nu_e}  <  2.5 
\eV$~\cite{MoscowH3} and 
$m_{\nu_e} <  2.9  \eV$~\cite{Mainz}. 
There are prospects to 
increase the sensitivity  
of the \hbeta experiments
to  $m_{\nu_e} \sim (0.3 - 1.0)$ eV 
\cite{KATRIN} (the KATRIN project).

   The problem of the nature of massive neutrinos
can be addressed in experiments studying processes 
in which  the total lepton charge $L$
is not conserved and changes by two units,
$\Delta L = 2$.
The process most sensitive to the 
existence of massive Majorana neutrinos 
(coupled to the electron)
is the neutrinoless double $\beta$ 
($\betabeta-$) decay of
certain even-even nuclei (see, e.g., \cite{BiPe87}): 
$(A,Z) \rightarrow (A,Z+2) + e^{-} + e^{-}$.
If the $\betabeta-$ decay is generated 
only by the 
left-handed  charged current weak interaction
through the
exchange of virtual light
massive Majorana neutrinos, 
the probability amplitude of this process 
is proportional 
to the ``effective Majorana mass parameter'':
\begin{equation}
\meff \equiv  \left| | U_{\mathrm{e} 1}|^2~ m_1 +
|U_{\mathrm{e} 2}|^2~e^{i \alpha_{21}} m_2 +
| U_{\mathrm{e} 3}|^2~e^{i \alpha_{31}} m_3 \right|,
\label{meff}
\end{equation}

\noindent where $m_j$ is the mass of the 
Majorana neutrino $\nu_j$,  
$U_{\mathrm{e}j}$
is the element of the 
Pontecorvo-Maki-Nakagawa-Sakata (PMNS)
neutrino (lepton) 
mixing matrix \cite{BP57,MNS62}, and $\alpha_{j1}$,
$j=2,3$, are two Majorana
CP-violating phases \cite{BHP80}.  
If  CP-parity is conserved we have \cite{LW81,BNP84}
 $\alpha_{j1} =k \pi$, $k=0,1,2...$.
  Many experiments are searching  for 
$\betabeta-$decay. 
No  indications that this process takes place were found so far.
A most stringent constraint on the 
value of \meff was obtained in the $^{76}$Ge 
Heidelberg-Moscow 
experiment \cite{76Ge00} 
$ \meff < 0.35 \ \mathrm{eV}$ ($90\%$ C.L.).
The  IGEX collaboration has obtained \cite{IGEX00}
$\meff < (0.33 \div 1.35) \ \mathrm{eV}$ 
( $90\%$ C.L.).
  Higher sensitivity to the value of 
$\meff$ is planned to be 
reached in several $\betabeta-$decay experiments
of a new generation \cite{NEMO3}. 
The NEMO3  and the planned CUORE
experiments aim to achieve 
a sensitivity to values of $\meff \cong 0.1~$eV.
A sensitivity to $\meff \cong 10^{-2}~$eV,
is planned to be reached
in the GENIUS and  EXO experiments.
 
The present article represents a continuation
of the studies of the physical implications of
the possible future results on
\meff and on $m_{\nu_e}$
performed in Refs. \cite{BPP1,BPP2,PPW}.
The article is based on the
work done in \cite{PPW}.
We generalize the results, derived in \cite{PPW}
by using the best fit values of the 
input neutrino oscillation
parameters, to the case when some of 
these parameters take values
in their 90\% C.L. allowed regions.
Earlier studies on the subject were performed, e.g.,
in Refs. \cite{bbpapers1,bbpapers2,bbpapers3}.
Recent relevant studies include Ref. \cite{Czakon01}.
\vspace{-0.1cm}
\section{Constraining or Determining the Lightest Neutrino Mass 
$m_1$ and/or the Majorana CP-Violating Phases}
\vspace{-0.2cm}

  In this Section, assuming $3-\nu$ mixing, 
we discuss the 
information that future 
\betabeta-decay and/or $^3$H $\beta-$decay 
experiments can provide on the lightest neutrino mass 
$m_1$ and on the CP-violation generated by the 
Majorana CP-violating phases 
$\alpha_{21}$ and  $\alpha_{31}$.
We number the three neutrinos 
with definite mass $\nu_j$ in 
such a way that their masses obey 
$m_1 < m_2 < m_3$. The two cases of 
neutrino mass spectrum are analyzed:
spectrum with normal hierarchy,
$\deltasol \equiv \deltadue$, 
and with inverted hierarchy,
$\deltasol \equiv \deltatre$,
$\deltasol > 0$ being the neutrino mass
squared difference characterizing
the solar neutrino oscillations
(see, e.g., Refs. \cite{BPP1,PPW}).
In both cases we use in the 
analysis which follows
fixed values of the mixing parameters
$|U_{\mathrm{e}3}|^2$ (or $| U_{\mathrm{e}1}|^2$) and of the angle
$\theta_{\odot}$, which are 
constrained by the CHOOZ and   
the solar neutrino data~\footnote{
In our further discussion we assume
$\cos2\theta_{\odot} \geq 0$,
which is favored by the analyses 
of the solar neutrino data 
\cite{ConchaSNO}.
The modification of the relevant formulae
and of the results 
in the case $\cos2\theta_{\odot} < 0$ 
is rather straightforward.}, respectively.
We allow  \deltasol and \deltaatm,
which characterizes the oscillations 
of atmospheric neutrinos,
to vary within their 90\% C.L.
allowed intervals found in the analyses 
of the solar and atmospheric neutrino data
in Refs. \cite{Gonza3nu,ConchaSNO}.
We denote the minimal and maximal values 
in these intervals by 
$\deltasolmin$, $\deltasolmax$,
$\deltaatmmin$ and $\deltaatmmax$. 
The results thus obtained are summarized in Fig. 1
(normal neutrino mass hierarchy) and 
in Fig. 2 (inverted hierarchy).

\vspace{-0.1cm}
\subsection{Normal Mass Hierarchy: $\deltasol \equiv \deltadue$}
\vspace{-0.2cm}

  If $\deltasol = \Delta m^2_{21}$,
for any given solution 
of the solar neutrino problem 
LMA MSW, LOW-QVO, SMA MSW, 
\meff can lie anywhere
between 0 and the present upper limits,
 as Fig. 1 indicates.
This conclusion 
does not change even 
under the most favorable
conditions for the determination of \meff,
namely, even when \deltaatm, \deltasol,
$\theta_{\odot}$ and $\theta$ are known
with negligible uncertainty.
The further conclusions 
in the case of the 
LMA MSW solution of the solar neutrino problem,
which are illustrated
in Fig. 1, are now summarized.
  
{\bf Case A.} An experimental upper 
limit on \meff, $\meff < \meff_{exp}$,
will determine a maximal value of $m_1$,
 $m_1 < (m_1)_{max}$. The value
$(m_1)_{max}$ is fixed by one of the two equalities: 
\begin{align}
\label{maxm1nhLMA}
  \meff_{ \! exp} \! &=   
 \left| \!  \left( \!  m_1\cos^2  \!  \! \theta_\odot - \sqrt{m_1^2 + \deltasolmax \! }
\sin^2  \! \theta_\odot \!  \right) \!  (1 - \utremax  \! \! ) -
\sqrt{m_1^2 + \deltaatmmax} \utremax \!  \! 
\right|; \\
\label{maxm1nhLMAcos0}
 \meff_{ \! exp} \!  &= 
 \left| \!  \left( \!  m_1 \cos^2 \!  \!  \theta_\odot - \sqrt{m_1^2 + \deltasolmin \! }
\sin^2 \!  \theta_\odot \!  \right)  \! (1 - \utremin \! \!  ) + 
\sqrt{m_1^2 + \deltaatmmax} \utremin \!  \! 
\right|.
\end{align}
%
\noindent 
Eq.~(\ref{maxm1nhLMA}) is valid for 
$\cos 2\theta_\odot > (\deltasol / m_1^2 ) \sin^4 \theta_\odot$,
otherwise
eq.~(\ref{maxm1nhLMAcos0}) should be used.

   For the quasi-degenerate  mass spectrum
one has $m_1 \gg \deltasol,\deltaatm$,  
$m_1 \cong m_2 \cong m_3 \cong m_{\nu_e}$, and~\cite{PPW,Czakon01} 
\begin{equation}
 (m_1)_{max} \cong \frac{\meff_{exp}}
{\left|\cos 2\theta_\odot (1 - \utremax) 
- \utremax \right|}~.
\label{maxm1nhLMAQD}
\end{equation}
%
\noindent 
If $ |\cos 2\theta_\odot (1 - \utremax) 
- \utremax |$ is sufficiently small,
the upper limit on $m_{\nu_e}$ obtained in 
$^3$H $\beta-$decay experiments
could yield  a more 
stringent upper bound  on $m_1$ 
than the bound  from  
the limit on \meff.

{\bf Case B. } 
A measurement  of $\meff = (\meff)_{exp}
\gtap 0.02$ eV
would imply that $m_1 \gtap 0.02$ eV and 
thus a  mass spectrum with partial hierarchy
or of quasi-degenerate type \cite{BPP1}. 
The lightest neutrino mass
will be constrained to lie in the interval,
$(m_1)_{min} \leq m_1 \leq (m_1)_{max}$, 
where $(m_1)_{max}$ and $(m_1)_{min}$
are determined respectively by 
eq. (\ref{maxm1nhLMA}) or (\ref{maxm1nhLMAcos0})
and by the equation~\cite{PPW}:
\begin{equation}
(m_1\cos^2 \!  \theta_\odot + \sqrt{m_1^2 + \deltasolmax \! }
\sin^2 \!  \theta_\odot) (1 - \utremax \!  \! ) \!  + \! 
\sqrt{m_1^2 + \deltaatmmax \! } \utremax \!  \!  \! 
= \meff_{ \! exp}.
\label{minm1nhLMA01}
\end{equation}
%
\noindent 
The limiting values of $m_1$ 
correspond to the case of CP-conservation.
If it holds that  $\deltasol \ll m_1^2$, (i.e., for 
$\deltasol \ltap 10^{-4}~{\rm eV^2}$),
$(m_1)_{min}$ to a good approximation
is independent of $\theta_{\odot}$,
and for $\sqrt{\deltaatm} |U_{\mathrm{e}3}|^2 \ll m_1$,
which takes place in the case we consider 
as $|U_{\mathrm{e}3}|^2 \ltap 0.05$,
we have $(m_1)_{min} \cong (\meff)_{exp}$.
For $|U_{\mathrm{e}3}|^2 \ll \cos2\theta_{\odot}$,
which is realized 
in Fig. 1 for $|U_{\mathrm{e}3}|^2 \ltap 0.01$, 
practically all the region between
$(m_1)_{min}$ and $(m_1)_{max}$,
$(m_1)_{min} < m_1 < (m_1)_{max}$,
corresponds to violation of the CP-symmetry.
If $|U_{\mathrm{e}3}|^2$ is non-negligible
with respect to $\cos2\theta_{\odot}$,
e.g., if $|U_{\mathrm{e}3}|^2 \cong (0.02 - 0.05)$
for the values of $\cos2\theta_{\odot}$
used to derive the right panels in Fig. 1,  
one can have 
$(m_1)_{min} < m_1 < (m_1)_{max}$
if the  CP-symmetry is violated, as well as  
in  two specific cases of CP-conservation~\cite{PPW}.
One of these two CP-conserving values of $m_1$,
corresponding to $\eta_{21} = - \eta_{31} = -1$,
can differ considerably from the two limiting values
(see Fig. 1). 
In general, the knowledge of the value
of \meff alone will not allow to 
distinguish the case
of CP-conservation 
from that of CP-violation.

{\bf Case C.} 
It might be possible to 
determine whether
CP-violation due to the Majorana 
phases takes place in the lepton sector if 
both \meff and $m_{\nu_e}$ are measured.
Since prospective measurements 
are limited to $(m_{\nu_e})_{exp} \gtap 0.35$ eV,
the relevant neutrino mass spectrum is of 
quasi-degenerate type, 
$m_1 \cong m_2 \cong m_3 \cong m_{\nu_e}$ (see, e.g., \cite{BPP1}) 
and one has
$m_1 > 0.35$ eV.
If we can neglect $|U_{\mathrm{e}3}|^2$ 
 (i.e., 
if $\cos 2\theta_\odot \gg |U_{\mathrm{e}3}|^2$),
a value of $m_{\nu_e} \cong m_1$, satisfying
$(m_1)_{min} < m_{\nu_e} < (m_1)_{max}$,
where $(m_1)_{min}$ and $(m_1)_{max}$ are determined by
eqs. (\ref{minm1nhLMA01}) and (\ref{maxm1nhLMA})
or (\ref{maxm1nhLMAcos0}),
would imply that the CP-symmetry 
does not hold in the lepton sector. 
In this case one would obtain 
correlated constraints
on the CP-violating phases 
$\alpha_{21}$ and $\alpha_{31}$ 
\cite{BPP1,Rodej00}.
This appears to be the only possibility 
for demonstrating   
CP-violation  due to Majorana  
CP-violating phases in the case of
$\deltasol \equiv \deltadue$ under discussion~\cite{PPW}.
In order to reach a definite conclusion
concerning CP-violation  due to the Majorana  
CP-violating phases, considerable accuracy
in the measured values of
\meff and $m_{\nu_e}$ is required.
For example, if the oscillation experiments
give the result $\cos2\theta_{\odot} \leq 0.3$
and $\meff = 0.3$ eV, a value of 
$m_{\nu_e}$  between 0.3 eV and 1.0 eV
would demonstrate CP-violation.
This requires better
than 30\% accuracy on both measurements.
The accuracy requirements become
less stringent if the upper limit on 
$\cos2\theta_{\odot}$ is smaller. 
 
  If $\cos2\theta_{\odot} > |U_{\mathrm{e}3}|^2$ but 
$|U_{\mathrm{e}3}|^2$ cannot be neglected
in \meff,  
there exist two 
CP-conserving values of  $m_{\nu_e}$
in the interval $(m_1)_{min} < m_{\nu_e} < (m_1)_{max}$~\cite{PPW}.
The one that can significantly differ 
from the extreme values of the interval
corresponds to the specific case of 
CP-conservation with 
$\eta_{21} = - \eta_{31} = -1$ (Fig. 1).

{\bf Case D. } 
A measured value of $m_{\nu_e}$,
$(m_{\nu_e})_{exp} \gtap 0.35$ eV, satisfying 
$(m_{\nu_e})_{exp} > (m_1)_{max}$,
where $(m_1)_{max}$ is determined from the upper limit
on \meff, eq. (\ref{maxm1nhLMA}) or (\ref{maxm1nhLMAcos0}), 
in the case
the \betabeta-decay is not 
observed, might imply 
that the massive neutrinos are Dirac particles.
If \betabeta-decay has been observed and \meff
measured, the inequality
$(m_{\nu_e})_{exp} > (m_1)_{max}$, 
would lead to the conclusion that
there exist contribution(s) to
the \betabeta-decay rate other than 
due to the light Majorana neutrino exchange
(see, e.g., \cite{bb0nunmi} 
and the references quoted therein) that 
partially cancel the
one from the 
Majorana neutrino exchange.

 A measured value of \meff, 
$( \meff)_{\mathrm{exp}} \gtap 0.01 \  \mathrm{eV}$,
and a  measured value of 
$m_{\nu_e}$ or an  upper bound on $m_{\nu_e}$
such that $m_{\nu_e} < (m_1)_{min}$,
where $(m_1)_{min}$ is determined by 
eq. (\ref{minm1nhLMA01}), 
would imply that there are contributions to the
\betabeta-decay rate in addition to the ones
due to the light Majorana neutrino exchange
(see, e.g., \cite{bb0nunmi}), 
which enhance the \betabeta-decay rate
and signal the existence of new $\Delta L =2$
processes beyond those induced 
by the light Majorana neutrino exchange
in the case of left-handed charged current  weak 
interaction.

{\bf Case E.} 
An actual measurement of 
$\meff \ltap 10^{-2}$ eV is unlikely, 
but it is illustrated in Fig. 1  to show
the interpretation of such a result. 
There always remains an upper limit on $m_1$,
 $(m_1)_{max}$, determined by
eq.~(\ref{maxm1nhLMA}) or by eq.~(\ref{maxm1nhLMAcos0}).
For $\utre^2 \geq \utretilda$,
where $\utretilda \equiv (\sqrt{\deltasolmin} \sin^2 \theta_\odot ) 
/ \sqrt{\deltaatmmax}$,
the cancellations between the different 
terms  allow to have $(m_1)_{min}=0$.
Both the cases of CP-conservation and 
CP-violation are allowed.
In principle, there exists the possibility 
to have regions of the parameter space 
where only the case of CP-violation is allowed.
However, in order to
 establish that 
\meff and  $m_1$  lie in those regions, 
they must be known with a precision
which is far beyond that aimed at in the currently
planned future experiments.

If $\utre^2 < \utretilda$, 
and \meff lies in the interval
$\meff_{-}~ \leq \meff \leq \meff_{+}$,
defined  by
\begin{align} 
\meff_{+} &= |\sqrt{\deltasolmax}
(\sin^2 \theta_\odot)_{\mbox{}_\mathrm{MAX}} (1 - \utremax) +
\sqrt{\deltaatmmax} \utremax|, \\
\meff_{-} &= |\sqrt{\deltasolmin }
(\sin^2 \theta_\odot)_{\mbox{}_\mathrm{MIN}} (1 - \utremax) -
\sqrt{\deltaatmmax} \utremax|, 
\end{align}
where $(\sin^2 \theta_\odot)_{\mbox{}_\mathrm{MIN}}$
and $(\sin^2 \theta_\odot)_{\mbox{}_\mathrm{MAX}}$ 
are respectively the minimal and maximal allowed values 
of $\sin^2 \theta_\odot$
 in the LMA solution region,
the lower limit on $m_1$ goes to zero. 
All the CP-parity patterns in the case 
of CP-conservation as well as the violation of
the CP-symmetry
are possible.
If $\utre^2 < \utretilda$ and  $\meff < \meff_{-}~$,
 $(m_1)_{min}$ is determined by the following equation:
\begin{equation}
    \meff_{exp}  = \left| (m_1\cos^2 \theta_\odot - \sqrt{m_1^2 + \deltasolmin}
\sin^2 \theta_\odot) (1 - \utre^2) +
\sqrt{m_1^2 + \deltaatmmax} \utre^2
\right|. 
\label{minm1nhLMA}
\end{equation}
%
\noindent 
Under the above conditions, 
the case of CP-conservation corresponding to
 $\eta_{21} = \pm \eta_{31} = 1$ will be excluded.
At the same time both the case of  CP-conservation with
$\eta_{21} = \pm \eta_{31} = -  1$ and 
that of CP-violation will be allowed. 

It should be noted also that 
one can have $\meff = 0$ for $m_1 = 0$
in the case of CP-invariance if 
$\eta_{21} = - \eta_{31}$
and the relation
$\sqrt{\deltasol}
\sin^2 \theta_\odot (1 - |U_{\mathrm{e}3}|^2) =
\sqrt{\deltaatm} |U_{\mathrm{e}3}|^2$ holds.
Finally, there would seem to be no practical possibility
to determine the Majorana CP-violating phases.

   The analysis of the {\it Cases A - E} 
for the LOW-QVO solution of the     
solar neutrino problem leads to the same
qualitative conclusions as those obtained above
for the LMA MSW solution. 
The conclusions differ,
however, in the case of the SMA MSW solution 
and we will discuss them next briefly~\cite{PPW}.
An experimental upper limit on 
\meff ({\it Case A}) in the range
$\meff_{exp} \geq 10^{-2}$ eV, would imply 
in the case of the SMA MSW solution, 
$m_1 < \meff_{exp}(1 - 2 \utremax)^{-1}$.
For values of $\meff \gtap 10^{-2}$ eV,
the maximum and minimum values of 
$m_1$ are extremely close:
$(m_1)_{min} \cong \meff_{exp}$.
As a result, 
a measurement of \meff ({\it Case B}) 
practically determines $m_1$,
$m_1 \cong \meff$. However,  
no information about CP-violation
generated by the Majorana phases 
can be obtained by the measurement of
\meff  (or of \meff and $m_{\nu_e}$) 
\cite{BPP1}. If both 
$\meff \gtap 0.02$ eV and 
$m_{\nu_e}\gtap 0.35$ eV 
would be measured ({\it Case C}),
the relation $m_1 \cong (\meff)_{exp} 
\cong (m_{\nu_e})_{exp}$
should hold. The conclusions in the 
{\it Cases D} and {\it E}
are qualitatively the same as for the LMA MSW 
solution.
 
\vspace{-0.1cm}
\subsection{Inverted Mass Hierarchy: $\deltasol \equiv \deltatre$}
\vspace{-0.2cm}
  Consider next the possibility of a 
neutrino mass spectrum with inverted 
hierarchy, which is illustrated in Fig. 2. 
A comparison of Fig. 1 and Fig. 2
reveals two major differences 
in the predictions for \meff:
if $\deltasol \equiv \deltatre$, 
i) even in the case of $m_1 \ll m_2 \cong m_3$
(i.e., even if $m_1 \ll 0.02$ eV), \meff can 
exceed $\sim 10^{-2}$ eV
and can reach the value of $\sim 0.08$ eV 
\cite{BPP1}, and 
 ii) a more precise determination of 
\deltaatm, \deltasol, 
$\theta_{\odot}$ 
and $\sin^2\theta = |U_{e1}|^2$, 
can lead to a lower limit on the possible values of 
\meff  \cite{BPP1}.
For the LMA and the LOW-QVO
solutions, ${\rm min}(\meff)$
will depend, in particular,
on whether CP-invariance 
holds or not in the lepton sector, 
and if it holds -
on the relative CP-parities 
of the massive Majorana neutrinos.
All these possibilities are parametrized
by the values of the two  
CP-violating phases,
$\alpha_{21}$ and $\alpha_{31}$,
entering into the expression for \meff.
The existence of a significant lower
limit on the possible values
of \meff depends crucially in the cases of the
LMA and LOW-QVO solutions on the
minimal value of $|\cos2\theta_{\odot}|$,
$|\cos2\theta_{\odot}|_{\mbox{}_\mathrm{MIN}}$, 
allowed by the data. Up to corrections
$\sim 5\times 10^{-3}$ eV, the minimum
value of \meff is for these
two solutions (see, e.g., \cite{BPP1}):
\begin{equation}
(\meff)_{\mbox{}_\mathrm{MIN}} \simeq
\left| \sqrt{\deltaatmmin} | ( \cos 2 \theta_\odot)|_{\mbox{}_\mathrm{MIN}} 
(1 - \uunomax) \right|.
\label{minmeffLA}
\end{equation}

\noindent The min(\meff) in eq. (\ref{minmeffLA}) 
is reached in the case of CP-invariance and
$\eta_{21} = - \eta_{31} = \pm 1$. 

 We shall discuss next briefly 
the implications of the results
of future \betabeta-decay and 
\hbeta experiments.
We follow the same line of analysis 
we have used for neutrino mass spectrum 
with normal hierarchy. Consider the case 
of the LMA MSW solution 
of the solar neutrino problem. 

{\bf Case A.} An experimental upper 
limit on \meff, $\meff < \meff_{exp}$,
which is larger than the 
minimal value of \meff, 
$\meff_{min}^{ph}$, 
predicted by taking into account 
all uncertainties in the values of 
the relevant input parameters 
(\deltaatm, \deltasol, $\theta_{\odot}$, etc.),
$\meff_{exp} \geq \meff_{min}^{ph}$,
will imply an upper limit on $m_1$, $m_1 < (m_1)_{max}$.
The latter is determined by the equality: 
%
\begin{equation}
\left|  \!   \left( \!  \sqrt{m_1^2 + \deltaatmmin \! \! \! \!  - \deltasolmax \!}
 \! \cos^2 \theta_\odot - \sqrt{m_1^2 + \deltaatmmin \!}
 \! \sin^2 \theta_\odot \!  \right) \! (1 - |U_{\mathrm{e}1}|^2  \!)
\pm m_1 |U_{\mathrm{e}1}|^2 \right| \!   =  \! \meff_{ \! exp},
\label{maxm1ih}
\end{equation}
%
where the $+$ ($-$) corresponds to a negative
(positive) value of the expression in the big 
round brackets~\cite{PPW}.
  For the quasi-degenerate neutrino mass spectrum
($m_1 \gg \deltasol,\deltaatm$,  
$m_1 \cong m_2 \cong m_3 \cong m_{\nu_e}$), 
$(m_1)_{max}$ is given by eq. (\ref{maxm1nhLMAQD})
in which $|U_{\mathrm{e}3}|^2$ is replaced by 
$|U_{\mathrm{e}1}|^2$.
Correspondingly, the conclusion that if 
$|\cos 2\theta_\odot (1 - |U_{\mathrm{e}1}|^2) 
- |U_{\mathrm{e}1}|^2|$ is 
sufficiently small,
the upper limit on $m_1 \cong m_{\nu_e}$, 
obtained in $^3$H $\beta-$decay, can 
be more stringent than 
the upper bound  on $m_1$, 
implied by the limit on \meff,
remains valid.

   An experimental upper 
limit on \meff,
which is smaller than the 
minimal possible value of \meff, 
$\meff_{exp} < \meff_{min}^{ph}$,
would imply that either 
i) the neutrino mass spectrum is not
of the inverted hierarchy type, or ii) that 
there exist contributions to
the \betabeta-decay rate other than 
due to the light Majorana neutrino exchange
(see, e.g., \cite{bb0nunmi}) that 
partially cancel the
contribution from the 
Majorana neutrino exchange. 
The indicated result might also 
suggest that the massive neutrinos 
are Dirac particles.

{\bf Case B.} 
A measurement of $\meff = (\meff)_{exp}
\gtap \sqrt{\deltaatmmax}(1 -  \uunomin)
\cong (0.04  - 0.08)$ eV, where we have used the 
90\% C.L. allowed regions of \deltaatm and  
$|U_{\mathrm{e}1}|^2$ from \cite{Gonza3nu},
would imply the existence of a finite 
interval of possible values of $m_1$,
$(m_1)_{min} \leq m_1 \leq (m_1)_{max}$, 
with $(m_1)_{max}$ and $(m_1)_{min}$
given respectively by eq. (\ref{maxm1ih}) and 
 by the equality:
%
\begin{equation}
 m_1 \uunomin \! \! \!  \!  +  \! 
\left ( \!  \sqrt{m_1^2 + \deltaatmmax \! \! \! \! \!   - \deltasolmin \! \!  }
\cos^2 \theta_\odot 
 + \!   \sqrt{m_1^2 + \deltaatmmax \! \!  }
\sin^2 \theta_\odot \right )  \! (1 - \uunomin \! \!   ) 
= \meff_{ \! exp}.
\label{minm1ih}
\end{equation}
%
\noindent 
In this case $m_1 \gtap 0.04$ eV and 
the neutrino mass spectrum is with partial inverted 
hierarchy or of quasi-degenerate type \cite{BPP1}. 
The limiting values of $m_1$ 
correspond to CP-conservation.
For $\deltasol \ll m_1^2$, i.e.,
for $\deltasol \ltap 10^{-4}~{\rm eV^2}$,
$(m_1)_{min}$ is to a good approximation
independent of $\theta_{\odot}$ and
we have: \linebreak
 $\sqrt{((m_1)_{min})^2 + \deltaatmmax}~
(1 - \uunomin)\cong (\meff)_{exp}$.

   For negligible $|U_{\mathrm{e}1}|^2$
(i.e., $|U_{\mathrm{e}1}|^2 \ltap 0.01$ 
for the values of $\cos2\theta_{\odot}$ in Fig. 2),
essentially all of the interval between
$(m_1)_{min}$ and $(m_1)_{max}$,
$(m_1)_{min} < m_1 < (m_1)_{max}$,
corresponds to violation of the CP-symmetry.
If the term $\sim m_1 |U_{\mathrm{e}1}|^2$ cannot be neglected 
in eqs. (\ref{maxm1ih}) and (\ref{minm1ih}) 
(i.e., if $|U_{\mathrm{e}1}|^2 \cong (0.02 - 0.05)$
for the values of $\cos2\theta_{\odot}$ in Fig. 2),
there exists for a fixed $\meff_{exp}$
two CP-conserving values of $m_1$ in the  
indicated interval~\cite{PPW}, one of which differs
noticeably from the limiting values 
$(m_1)_{min}$ and  $(m_1)_{max}$
and corresponds to 
$\eta_{21} = - \eta_{31} = 1$ (Fig. 2).

 In general, measuring the value
of \meff alone will not allow to 
distinguish the case
of CP-conservation from that of CP-violation.
In principle, a measurement of 
$m_{\nu_e}$, or even an upper limit 
on $m_{\nu_e}$, smaller than $(m_1)_{max}$,
could be a signal of CP-violation,
as Fig.~2 (upper panels) shows.
However, unless $\cos2\theta_{\odot}$
is very small, the required values of 
$m_{\nu_e}$ are less than
prospective measurements. For example,
as is seen in Fig. 2, middle left panel,
for $\cos2\theta_{\odot} = 0.1$ and
$\meff = 0.03$ eV, one needs to find
$m_{\nu_e} < 0.35$ eV to demonstrate
CP-violation.

  If the measured value of \meff lies 
in the interval
$(\meff_{-})_{\mbox{}_\mathrm{MAX}} \leq \meff
 \leq (\meff_{+})_{\mbox{}_\mathrm{MIN}}$, where
\vspace{-0.2cm}
%
\begin{align}
\vspace{-0.2cm}
(\meff_{+})_{\mbox{}_\mathrm{MAX}} = 
| \sqrt{\deltaatmmax - \deltasolmin}
\cos^2 \theta_\odot + \sqrt{\deltaatmmax}
\sin^2 \theta_\odot | (1 - \uunomin), \\
\vspace{-0.2cm}
(\meff_{-})_{\mbox{}_\mathrm{MIN}} = 
| \sqrt{\deltaatmmin - \deltasolmax}
\cos^2 \theta_\odot - \sqrt{\deltaatmmin}
\sin^2 \theta_\odot | (1 - \uunomax), 
\label{0m1ih}
\end{align}
%
\noindent 
we would have $(m_1)_{min} = 0$.
Furthermore, if one finds that 
$\meff < (\meff_{+})_{\mbox{}_\mathrm{MIN}}$ with
$  (\meff_{+})_{\mbox{}_\mathrm{MIN}} = 
| \sqrt{\deltaatmmin - \deltasolmax}
\cos^2 \theta_\odot +  $     
$ \sqrt{\deltaatmmin}
\sin^2 \theta_\odot | (1 - \uunomax)$,  \! 
 the case of CP-conservation corresponding to
$ \eta_{21} = \eta_{31} = \pm 1$
will be excluded.
If $(\meff_{-})_{\mbox{}_\mathrm{MAX}} 
<  \meff \!  \! < \! (\meff_{\! +})_{\mbox{}_\mathrm{MIN}}$,
where $(\meff_{\! -})_{\mbox{}_\mathrm{MAX}} = 
\left| \!  \sqrt{\deltaatmmax \! \!  \!   - \deltasolmin \! } \! 
\cos^2 \!  \theta_\odot - \sqrt{\deltaatmmax \!  \! }
\sin^2 \theta_\odot \right|$ $ (1 - \uunomin)$ 
and $0 < m_1 < (m_1)_{\mathrm{max}}$,  
one will conclude  that the CP-symmetry is violated.

  {\bf Cases C.} As Fig. 2 indicates,
the discussions 
and conclusions are identical to 
the ones  in the same
cases for the neutrino mass spectrum 
with normal hierarchy, except that
 $(m_1)_{max}$ and $(m_1)_{min}$
are determined by eqs. (\ref{maxm1ih})
and (\ref{minm1ih}), and  
$|U_{\mathrm{e}3}|^2$ must be substituted
by $|U_{\mathrm{e}1}|^2$ in the 
relevant parts of the analysis.

 {\bf Case D.}
 If $m_{\nu_e}$ is measured 
and  $(m_{\nu_e})_{exp} \gtap 0.35 \ \mathrm{eV}$ 
but the \betabeta-decay is not observed 
or is observed and
$(m_{\nu_e})_{exp} > (m_1)_{max}$,
where $(m_1)_{max}$ is determined by eq. (\ref{maxm1ih}),
the same considerations and conclusions as
 in the Case D for the normal hierarchy
mass spectrum apply.

A measured value of \meff, 
$( \meff)_{\mathrm{exp}} \gtap 0.1 \ \mathrm{eV}$,
in the case when the measured value of 
$m_{\nu_e}$ or the upper bound on $m_{\nu_e}$
are such that $m_{\nu_e} < (m_1)_{min}$,
where  $(m_1)_{min}$ is determined by 
eq. (\ref{minm1ih}), 
would lead to the 
same conclusions as  in the Case D
for the normal hierarchy
mass spectrum.

  {\bf Case E.} 
It is possible to have a 
measured value of 
$\meff \ltap 10^{-2}$ eV 
in the case of 
the LMA MSW solution and 
neutrino mass spectrum
with inverted hierarchy 
under discussion only if 
$\cos2\theta_{\odot}$ is 
rather small, $\cos2\theta_{\odot}\ltap 0.2$.
A measured value of 
$\meff < \meff_{min}^{ph}$ would imply
that either
the neutrino mass spectrum is not
of the inverted hierarchy type, or  that 
there exist contributions to
the \betabeta-decay rate other than 
due to the light Majorana-$\nu$ exchange
that partially cancel the
contribution from the 
Majorana-$\nu$ exchange.

   The above conclusions hold with minor 
modifications (essentially of the numerical 
values involved) for the 
LOW-QVO solution as well. In the case of the 
SMA MSW solution we have, as is well-known, 
$\sin^2\theta_{\odot} << 1$ and 
$\deltasol \ltap 10^{-5}~{\rm eV^2}$ (see, e.g.,
\cite{ConchaSNO}). Consequently, the analog
of eq. (\ref{maxm1nhLMAQD}) in {\it Case A} reads
$(m_1)_{max} \cong \meff_{exp}
(1 - 2|U_{\mathrm{e}1}|^2)^{-1}$.
The conclusions in the  {\it Cases B - D}
are qualitatively the same 
as in the case of
neutrino mass spectrum with normal hierarchy.
In particular, a measured value of 
$\meff > \meff_{+} \cong \sqrt{\deltaatmmax}
(1 - \uunomin)$, would essentially
determine $m_1$, $m_1 \cong (\meff)_{exp}$.
No information about CP-violation
generated by the Majorana phases 
can be obtained by the measurement of
\meff, or of \meff and $m_{\nu_e}$.
If both $\meff$ and  $m_{\nu_e}\gtap 0.35$ eV 
are measured, the relation $m_1 \cong (\meff)_{exp} 
\cong (m_{\nu_e})_{exp}$
should hold.
If it is found that 
$\meff = \sqrt{\deltaatm} (1 - \uuno)$,
one would have 
$0 \leq m_1 \leq (m_1)_{max}$, where 
$(m_1)_{max}$ is determined by eq. (\ref{maxm1ih})
in which effectively  
$\sin^2\theta_{\odot} = 0$,
$\cos^2\theta_{\odot} = 1$,
and $\deltasol = 0$.
Finally, a measured value of 
$\meff < (\meff_{-})_{\mathrm{MIN}} \cong (\meff_{+})_{\mathrm{MIN}}
 \cong \sqrt{\deltaatmmin} (1 - \uunomax)$
would either indicate that 
there exist new additional contributions 
to the \betabeta-decay rate,
or that the SMA MSW solution
is not the correct solution of the
solar neutrino problem.

\vspace{-0.1cm}
\section{Conclusions}
\vspace{-0.2cm}
If \betabeta-decay will be 
detected  by present or upcoming experiments,
we will conclude that neutrinos are massive
Majorana particles and that the total 
lepton charge $L$ is not conserved.
  The observation of the
\betabeta-decay with a rate
corresponding to 
$\meff \gtap 0.02~$eV,
which is in the range of sensitivity of the 
future \betabeta-decay experiments, 
can provide unique information 
on the type of neutrino mass spectrum
and on the absolute values of 
neutrino masses. 
With additional information on the value of 
neutrino masses from \hbeta experiments
or the type of neutrino mass spectrum, 
one could obtain also information 
on the CP-violation in the lepton sector,
and - if CP-invariance holds - on the 
relative CP-parities
of the massive Majorana neutrinos.
The possibility of establishing CP-nonconservation
requires high precision measurements.
Given the precision of the future planned
\betabeta-decay and \hbeta experiments, 
it holds for a limited range of the values 
of the parameters involved.
\vspace{-0.2cm}


\section{Acknowledgements}
\vspace{-0.2cm}
S. P. and S. T. P. would like to thank L. Wolfenstein
for stimulating discussions which are at the origin of this paper. 
S. P. is  grateful to the organizers of this Workshop, 
for having being invited to such an interesting 
meeting which took place in a nice and stimulating atmosphere.
The work of S.P. was partly supported by the
Marie Curie Fellowship of the European Community
program HUMAN POTENTIAL under contract number
HPMT-CT-2000-00096.

\begin{figure}
\begin{center}
\epsfig{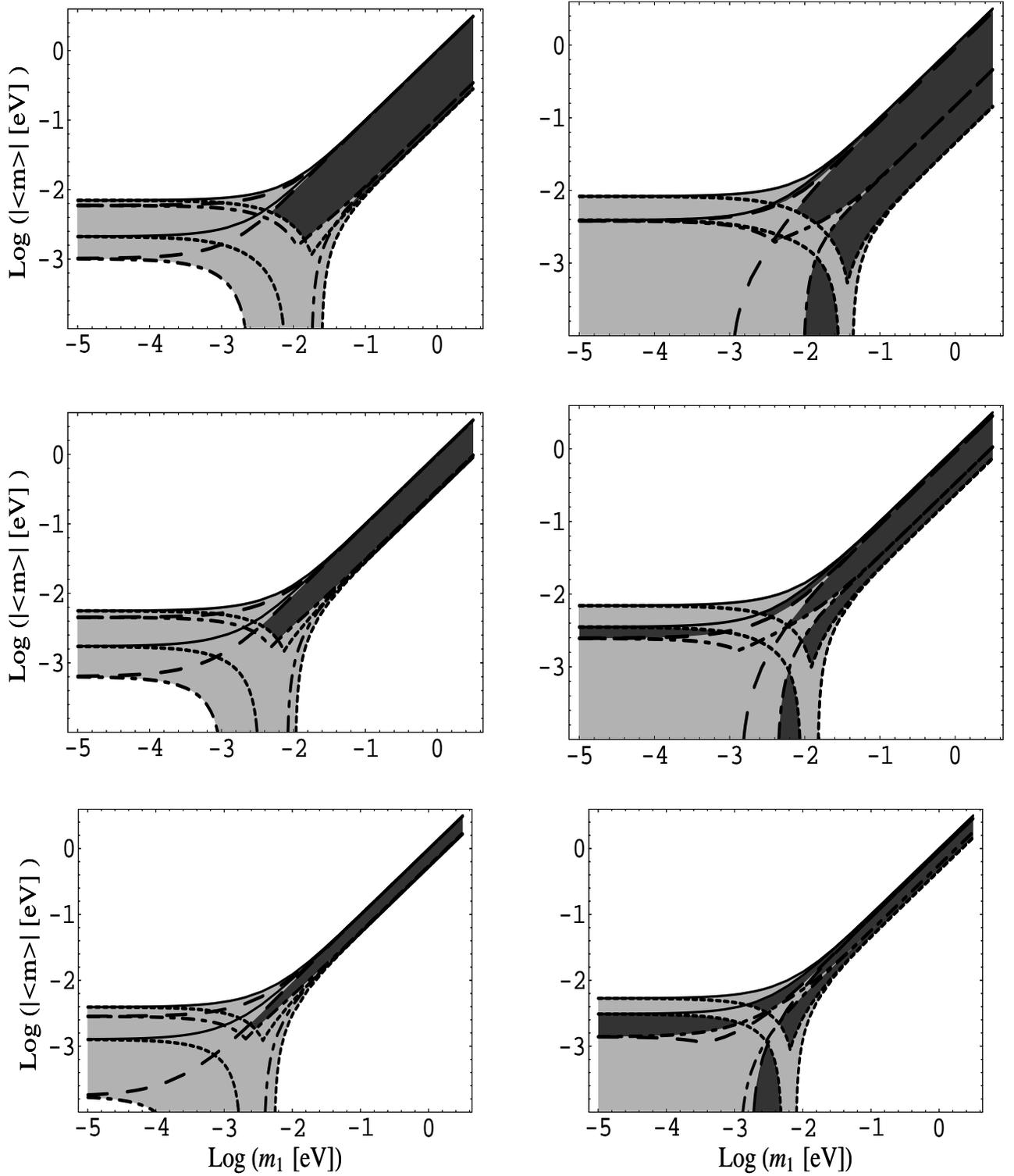}
\end{center}
\caption{
The dependence of \meff on $m_1$ 
for  $\deltasol = \Delta m_{21}^2$
in the case of 3-$\nu$ mixing and  of the LMA MSW solution
obtained at 90\%~C.L. in ref.\protect\cite{Gonza3nu},
for $\cos 2 \theta_\odot = 0.1$ (upper panels),
$\cos 2 \theta_\odot = 0.3$ (middle panels),
 $\cos 2 \theta_\odot = 0.54$ (lower panels), 
and for $|U_{\mathrm{e} 3}|^2 = 0.05$ 
(right panels)
and $|U_{\mathrm{e} 3}|^2 = 0.01$ 
(left panels).
The allowed values of \meff 
are constrained to lie
in the case of CP-conservation
{\it i)} in the medium-grey region 
between the two  thick solid lines if
$\eta_{21} = \eta_{31} = 1$,
{\it ii)} in the medium-grey region 
between the two long-dashed lines and the axes if
$\eta_{21} = - \eta_{31} = 1$,
{\it iii)} in the medium-grey region 
between the dash-dotted lines and the axes
if $\eta_{21} = - \eta_{31} = - 1$,
{\it iv)} in the medium-grey region 
between  the short-dashed lines 
if $\eta_{21} = \eta_{31} = - 1$.
In the case of CP-violation, the allowed region
for \meff covers all the grey region. 
Values of \meff in the dark grey region
signal CP-violation.} 
\label{fig:1}
\end{figure}

\pagebreak

\begin{figure}
\begin{center}
\epsfig{file=wfigproc02.epsi, height=20cm, width=17cm
}
\end{center}
\caption{The dependence of \meff on $m_1$ 
for $\deltasol = \Delta m_{32}^2$
in the case of 3-$\nu$ mixing
and of  the LMA MSW solution 
obtained at 90\%~C.L. in 
ref.\protect\cite{Gonza3nu},
for $\cos 2 \theta_\odot = 0$ (upper panels),
$\cos 2 \theta_\odot = 0.1$ (middle panels),
 $\cos 2 \theta_\odot = 0.3$ (lower panels), 
and for $|U_{\mathrm{e} 1}|^2 = 0.05$ 
(right panels)
and $|U_{\mathrm{e} 1}|^2 = 0.005$ 
(left panels).
The allowed regions
for \meff correspond to:
for $|U_{\mathrm{e} 1}|^2 = 0.005$
{\it i}) the medium-grey regions between the 
solid lines  
if $\eta_{21} = \eta_{31} = \pm 1$,
{\it ii}) the medium-grey regions between the dashed lines
(lowest left and middle left panels) 
or  the  dashed line (upper left panel)
if $\eta_{21} = - \eta_{31} = \pm 1$,
and all the grey regions 
if CP-invariance does not hold,  
and, for $|U_{\mathrm{e} 1}|^2 = 0.05$,
{\it iii}) the 
solid lines  
if $\eta_{21} = \eta_{31} =  1$,
{\it iv}) the 
long-dashed lines  
if $\eta_{21} = - \eta_{31} =   1$,
{\it v}) the 
dashed-dotted lines  
if $\eta_{21} = - \eta_{31} =  - 1$,
{\it vi}) the 
short-dashed lines  
if $\eta_{21} =  \eta_{31} = -  1$,
and all the grey regions 
if CP-invariance does not hold.
Values of \meff in the dark grey region
signal CP-violation.} 
\label{fig:2}
\end{figure}

\end{document}